\documentclass[final,5p,times,twocolumn,number,compress,longtitle]{elsarticle}

\usepackage{amsmath}
\usepackage{graphicx}
\usepackage{dcolumn}
\usepackage{bm}
\usepackage{booktabs}
\usepackage[T5,T1]{fontenc}
\usepackage{xspace}
\usepackage{siunitx}
\usepackage{float}
\usepackage{hyperref}
\hypersetup{colorlinks=true,allcolors=blue}
\usepackage{cleveref}
\usepackage{adjustbox}
\usepackage{multirow}

\newcommand{\cosz}{\ensuremath{\cos(\theta_{z})}\xspace}
\newcommand{\numu}{\ensuremath{\nu_\mu}\xspace}
\newcommand{\numubar}{\ensuremath{\overline{\nu}_\mu}\xspace}
\newcommand{\nue}{\ensuremath{\nu_e}\xspace}

\newcommand{\nutau}{\ensuremath{\nu_\tau}\xspace}
\newcommand{\nutaubar}{\ensuremath{\overline{\nu}_\tau}\xspace}

\newcommand{\LH}{\mathcal{L}}
\newcommand{\Uefsq}{\ensuremath{|U_{e4}|^2}\xspace}
\newcommand{\Umufsq}{\ensuremath{|U_{\mu4}|^2}\xspace}
\newcommand{\Utaufsq}{\ensuremath{|U_{\tau4}|^2}\xspace}
\newcommand{\Dmqfo}{\ensuremath{\Delta m_{41}^2}\xspace}

\usepackage[all]{hypcap}

\usepackage[caption=false]{subfig}

\usepackage{color}

\journal{Physics Letters B}

\begin{document}

\begin{frontmatter}

\title{Exploration of mass splitting and muon/tau mixing parameters for an eV-scale sterile neutrino with IceCube}

\author[loyola]{R. Abbasi}
\author[zeuthen]{M. Ackermann}
\author[christchurch]{J. Adams}
\author[madisonpac]{S. K. Agarwalla\fnref{india}}
\author[brusselslibre]{J. A. Aguilar}
\author[copenhagen]{M. Ahlers}
\author[dortmund]{J.M. Alameddine}
\author[bartol]{N. M. Amin}
\author[marquette]{K. Andeen}
\author[harvard]{C. Arg{\"u}elles}
% ...
\address[brusselslibre]{Universit{\'e} Libre de Bruxelles, Science Faculty CP230, B-1050 Brussels, Belgium}
% ...
\address[harvard]{Department of Physics and Laboratory for Particle Physics and Cosmology, Harvard University, Cambridge, MA 02138, USA}
% ...
\address[loyola]{Department of Physics, Loyola University Chicago, Chicago, IL 60660, USA}
\address[christchurch]{Dept. of Physics and Astronomy, University of Canterbury, Private Bag 4800, Christchurch, New Zealand}
% ...
\address[copenhagen]{Niels Bohr Institute, University of Copenhagen, DK-2100 Copenhagen, Denmark}
\address[dortmund]{Dept. of Physics, TU Dortmund University, D-44221 Dortmund, Germany}
% ...
\address[madisonpac]{Dept. of Physics and Wisconsin IceCube Particle Astrophysics Center, University of Wisconsin{\textemdash}Madison, Madison, WI 53706, USA}
% ...
\address[marquette]{Department of Physics, Marquette University, Milwaukee, WI 53201, USA}
% ...
\address[bartol]{Bartol Research Institute and Dept. of Physics and Astronomy, University of Delaware, Newark, DE 19716, USA}
% ...
\address[zeuthen]{Deutsches Elektronen-Synchrotron DESY, Platanenallee 6, D-15738 Zeuthen, Germany}
\fntext[india]{also at Institute of Physics, Sachivalaya Marg, Sainik School Post, Bhubaneswar 751005, India}
\fntext[chalmers]{also at Department of Space, Earth and Environment, Chalmers University of Technology, 412 96 Gothenburg, Sweden}
\fntext[tokyofn]{also at Earthquake Research Institute, University of Tokyo, Bunkyo, Tokyo 113-0032, Japan}
\fntext[caltech]{now at Division of Physics,
Mathematics and Astronomy, California Institute of Technology, Pasadena, CA, 91125, USA}

\begin{abstract}

We present the first three-parameter fit to a 3+1 sterile neutrino model using 7.634 years of data from the IceCube Neutrino Observatory on $\nu_\mu+\overline{\nu}_\mu$ charged-current interactions in the energy range 500--9976 GeV.
Our analysis is sensitive to the mass-squared splitting between the heaviest and lightest mass state ($\Delta m_{41}^2$), the mixing matrix element connecting muon flavor to the fourth mass state ($|U_{\mu4}|^2$), and the element connecting tau flavor to the fourth mass state ($|U_{\tau4}|^2$). 
Predicted propagation effects in matter enhance the signature through a resonance as atmospheric neutrinos from the Northern Hemisphere traverse the Earth to the IceCube detector at the South Pole. 
The remaining sterile neutrino matrix elements are left fixed, with $|U_{e4}|^2= 0$ and $\delta_{14}=0$, as they have a negligible effect, and $\delta_{24}=\pi$ is set to give the most conservative limits. 
The result is consistent with the no-sterile neutrino hypothesis with a probability of 4.3\%. 
Profiling the likelihood of each parameter yields the 90\% confidence levels: $ 2.4\,\mathrm{eV}^{2} < \Delta m_{41}^2 <9.6\,\mathrm{eV}^{2} $ , $0.0081 < |U_{\mu4}|^2 < 0.10$ , and $|U_{\tau4}|^2< 0.035$, which narrows the allowed parameter-space for $|U_{\tau4}|^2$.  However, the primary result of this analysis is the first map of the 3+1 parameter space exploring the interdependence of $\Delta m_{41}^2$, $|U_{\mu4}|^2$, and $|U_{\tau4}|^2$.

\end{abstract}

\begin{keyword}
atmospheric neutrino IceCube TeV oscillation sterile
\end{keyword}

\end{frontmatter}

\section{Introduction \label{sec:introduction}}

 Over the past quarter century, anomalies observed in short-baseline neutrino experiments have been studied within the context of vacuum neutrino oscillations involving the three known flavors ($\nu_e$, $\nu_\mu$ and $\nu_\tau$) and an additional hypothetical `sterile neutrino' ($\nu_s$) that has no left-handed weak interactions.
This ``3+1'' model predicts three related signatures: $\numu \to \nue$ appearance, \nue disappearance, and \numu disappearance.
While supporting evidence for anomalous oscillations have been found in short-baseline $\numu \to \nue$ appearance \cite{Athanassopoulos:1997pv,MiniBooNE:2020pnu} and \nue disappearance \cite{Barinov:2021asz,Mention:2011rk} experiments, no evidence has been found for $\nu_\mu$ disappearance \cite{MINOS:2017cae,SciBooNE:2011qyf}.
Further, other $\numu \to \nue$  appearance \cite{KARMEN:2002zcm,MicroBooNE:2021tya} and \nue disappearance \cite{PROSPECT:2020sxr,RENO:2020hva,STEREO:2019ztb} experiments fail to observe anomalous oscillations, and inconsistencies between all datasets lead to tensions within the ``3+1'' model \cite{Dentler:2018sju, Diaz:2019fwt,Gariazzo:2017fdh,Hardin:2022muu}.

Neutrino telescopes offer an alternative method to probe for the $\nu_\mu$ disappearance signal using atmospheric muon-flavor neutrino interactions at TeV-scale energies.
For these events, theoretical studies of 3+1 models predict, along with vacuum oscillations, a resonance signature induced by matter propagation effects on the $\bar \nu_\mu$ flux traversing the Earth which cause a deficit of ``up-going'' events~\cite{Akhmedov:1988kd,Krastev:1989ix,Chizhov:1998ug,Chizhov:1999az,Akhmedov:1999va,Nunokawa:2003ep}.
While the resonance signature will appear in the \numubar flux, current gigaton-scale high energy ($ {\gtrsim} \SI{1}{\TeV}$) neutrino telescopes cannot efficiently differentiate between a \numu and \numubar interaction, and so measure the combined ($\numu + \numubar$) rate. 
Further, due to sparse instrumentation, the oscillation signature is highly smeared in reconstructed energy.
However, the matter resonance effect may be sufficiently strong that neutrino telescopes still have the sensitivity to explore a wide range of 3+1 model parameter space.
Among existing neutrino telescopes, the IceCube Neutrino Observatory, a cubic-kilometer neutrino detector buried 1.5--2.5~km beneath the South Pole in the Antarctic glacier~\cite{Aartsen:2016nxy}, is both the largest and longest running experiment.
Therefore, it is the most promising in this search for new physics.

\section{Muon flavor disappearance in IceCube and the 3+1 model}

The 3+1 model parameterization introduces a new singlet state $\nu_s$ and a new mass state $\nu_4$.  The flavor mixing is described by expanding the PMNS 3-flavor mixing matrix to a $4\times 4$ mixing matrix $U_{3+1}$ that connects the flavor states ($\nu_e$, $\nu_\mu$, $\nu_\tau$, and $\nu_s$) to the mass states ($\nu_1$, $\nu_2$, $\nu_3$, and $\nu_4$):
\begin{equation}
U_{3+1} = \begin{pmatrix}
        U_{e 1} & U_{e 2} & U_{e 3} & U_{e 4}\\
        U_{\mu 1} & U_{\mu 2} & U_{\mu 3} & U_{\mu 4}\\
        U_{\tau 1} & U_{\tau 2} & U_{\tau 3 }& U_{\tau 4} \\
        U_{s 1} & U_{s 2} & U_{s 3} & U_{s 4}
\end{pmatrix}. \label{4mixmx}
\end{equation}
The matrix elements can be parameterized in terms of three new mixing angles ($\theta_{14}$, $\theta_{24}$ and $\theta_{34}$) and two new $CP$-violating parameters ($\delta_{14}$ and $\delta_{24}$). 
Within this multi-parameter model, three parameters, \Umufsq, \Utaufsq, and $\Delta m^2_{41}$, will most affect IceCube's TeV-scale muon-neutrino disappearance search.  This paper focuses on those parameters and sets the other parameters to $\Uefsq= 0$, $\delta_{14}=0$, and $\delta_{24}=\pi$. 

Setting $\Uefsq$ (or, equivalently, the mixing angle $\theta_{14}$) to zero is well-justified since the atmospheric neutrino flux in the TeV energy range has an electron-flavor content of less than 10\%~\cite{Fedynitch:2015zma,IceCube:2015mgt}.
As a result, contributions to muon flavor events involving \Uefsq, which would come mainly from $\nu_e \rightarrow \nu_\mu$ oscillations, can be safely neglected \cite{2011JHEP...07..084R,Esmaili:2012nz,Esmaili:2013vza}.
With $\Uefsq= 0$, the $CP$-violation term $\delta_{14}$ no longer contributes and is arbitrarily set to zero. 

In a matter-enhanced disappearance search, $\delta_{24}$ is expected to produce a minimal effect in the analysis for two reasons.
First,  at energies above \SI{500}{\GeV}, the lower bound of this analysis,   $\delta_{24}$ is sub-leading to the other parameters~\cite{Esmaili:2013vza}.
Second, non-zero $\delta_{24}$ weakens the disappearance of the \numu flux while simultaneously strengthening the disappearance of the \numubar flux, partially canceling out the effect to the overall \numu + \numubar flux.
However, since the \numu cross-section is larger than the \numubar cross-section, the overall \numu + \numubar event rate will slightly increase (i.e., the disappearance will weaken) with increasing $\delta_{24}$.
Therefore, the most conservative choice is to fix $\delta_{24}$ to the value that produces the weakest oscillation effect, $\delta_{24}=\pi$, which our sensitivity studies confirm. 

With the above assumptions, IceCube TeV-scale muon-flavor disappearance analyses depend upon two mixing matrix elements related to two mixing angles:
\begin{align}
    \Umufsq &= \sin^2 \theta_{24}\\ 
    \Utaufsq &= \sin^2 \theta_{34} \cos^2 \theta_{24}. \label{Udefs}
\end{align}
Terms depending on \Utaufsq have a smaller effect than those that depend on \Umufsq.
Therefore, previous matter-enhanced IceCube analyses have assumed $\theta_{34}=\Utaufsq = 0$~\cite{IceCube:2016rnb,IceCube:2020phf,IceCube:2020tka}, reducing the model to only two parameters and improving the speed of fits to data. However, as the precision of the IceCube data set has been enhanced through increased statistics (a factor of 15 between Ref.~\cite{IceCube:2016rnb} and Refs.~\cite{IceCube:2020phf,IceCube:2020tka}) and better estimation of systematic uncertainties, the effect of a non-zero \Utaufsq must now be studied.

\begin{figure*}[tb!]
    \centering 
    \includegraphics[width=.99\linewidth]{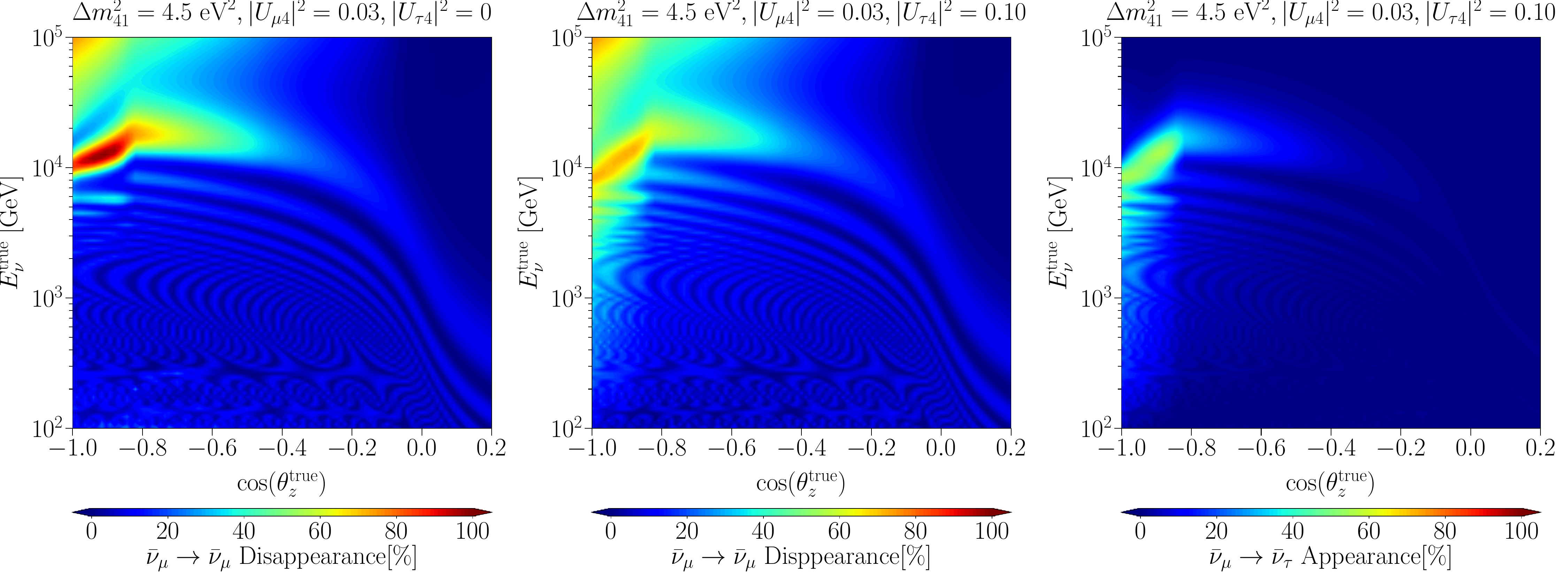}
    \caption{\textbf{Left:} This oscillogram shows $\numubar$ disappearance while propagating through the Earth, as a function of true neutrino energy and direction, assuming $\Dmqfo = \SI{4.5}{\eV\squared}$, $\Umufsq = 0.03$ and $\Utaufsq = 0.0$.
    \textbf{Center:} This oscillogram demonstrates how the $\numubar$ disappearance changes with the non-zero value of $\Utaufsq = 0.1$.
    \textbf{Right:} This oscillogram shows the $\numubar \to \nutaubar$ appearance once \Utaufsq has been set to the non-zero value.} 
    \label{fig:oscillograms}
\end{figure*}

\Cref{fig:oscillograms} presents `oscillograms' illustrating how the atmospheric neutrino flux changes for some representative sterile neutrino mixing parameters.
The left plot shows \numubar disappearance as a function of true neutrino energy and direction,
assuming a particular value of \Dmqfo and \Umufsq, while keeping $\Utaufsq = 0$.
The neutrino's direction serves as a proxy for the distance and amount of matter it traversed between creation and arrival at the IceCube detector.
The direction is given by \cosz, where $\theta_{z}$ is the zenith angle.
In these coordinates, $\cosz = -1$ corresponds to ``up-going'' events that traverse the Earth's diameter, while $\cosz = 0$ are events coming from the horizon.
For the chosen values of \Dmqfo and \Umufsq, the sterile-induced signature is the resonant deficit localized in direction and energy at $\cosz \approx \num{-1}$ and  $ E_{\nu} \sim \SI{e4}{\GeV}$.
These disappearing neutrinos primarily oscillate into the sterile $\nu_{s}$ state. 
The middle plot shows how the oscillogram changes further when introducing a non-zero $\Utaufsq$.
The \numubar resonant disappearance is smeared lower in energy and becomes more up-going.
In this scenario, there is also significant $\numubar \to \nutaubar$ appearance that further modifies the observed signal, as seen in the right plot. 

As seen in Table~\ref{tab:pastresults}, the present limits on \Utaufsq apply in restricted \Dmqfo ranges where rapid oscillations cause the experiments to be insensitive to this parameter, reducing the analyses to two-parameter fits in \Umufsq and \Utaufsq, which are correlated.
This motivates the present study to \emph{simultaneously} improve our knowledge of all three parameters, $\Dmqfo$, $\Umufsq$, and $\Utaufsq$.
This letter will present frequentist and Bayesian results that characterize these parameters.

\begin{table}[t] \centering
\begin{adjustbox}{width=\columnwidth}
\begin{tabular}{ c c c  } 
Experiment [Ref.] &  $\Utaufsq$ at 90\% (99\%) CL & $\Dmqfo$  \\ 
\hline
IceCube-DeepCore~\cite{IceCube:2017ivd} &  $< 0.15$ & > \SI{1}{\eV\squared} \\
Super-Kamiokande~\cite{Super-Kamiokande:2014ndf} & $< 0.18\ (0.23)$ & $>\SI{0.1}{\eV\squared}$ \\
T2K~\cite{T2K:2019efw}  & $ < 0.50$ & > \SI{0.1} {\eV\squared} \\
ANTARES~\cite{ANTARES:2018rtf} &  $< 0.40\ (0.68)$ & > $\SI{0.5}{\eV\squared}$ \\
 \end{tabular}
 \end{adjustbox}
\caption{Summary of previous limits on $\Utaufsq$ at the 90\% (99\%, if available) confidence level and the applicable $\Dmqfo$ range where the predicted oscillation is sufficiently fast to be effectively averaged over.  In these analyses there is a dependence between $\Utaufsq$ and $\Umufsq$ values, thus the applicable $\Umufsq$ parameter space for these constraints is also limited. \label{tab:pastresults}}  
\end{table}

\section{Event selection and analysis \label{sec:analysis}}

The analysis considers a muon deficit within the combined atmospheric and astrophysical muon-flavor neutrino and antineutrino fluxes.
The selection is dominated by the atmospheric flux that is initiated through cosmic ray interactions with the Earth's atmosphere.
To generate the atmospheric flux, IceCube uses the \texttt{MCEq} software~\cite{Fedynitch:2015zma}, which takes a specific cosmic ray model as input and produces the resulting flux, calculated according to a particular hadronic interaction model.
We use the Hillas-Gaisser 2012 H3a cosmic ray model~\cite{Gaisser:2011klf} with the SIBYLL 2.3c hadronic interaction model~\cite{Riehn:2017mfm}.
The flux further depends on the Earth's atmospheric density profile, obtained from data from the AIRS instrument on NASA's Aqua satellite~\cite{AIRS}.

The prompt atmospheric component originates from the decays of charmed mesons produced by cosmic ray interactions.   These heavier hadrons decay before substantial energy loss, resulting in a harder spectrum than the `conventional' flux discussed above.
We use the expected flux derived in Ref.~\cite{Bhattacharya:2015jpa} as input.
A prompt atmospheric component has yet to be identified in IceCube, with a 1 $\sigma$ upper limit found at a factor of \num{\sim 5} greater than the baseline model used~\cite{IceCube:2020wum}.
Within the energy range of interest for this analysis, the prompt component is expected to contribute less than 1\% to the overall event rate.

The astrophysical neutrino flux was first measured by IceCube~\cite{IceCube:2013low,IceCube:2014stg,IceCube:2020wum}, and is fit by an isotropic, falling power-law spectrum:
\begin{equation}
    \frac{dN_\nu}{dE} = \Phi_{\textrm{astro}} \times \left( \frac{E_\nu}{\SI{100}{\TeV}} \right)^{-\gamma_{\textrm{astro}}},
    \label{eq:astroflux}
\end{equation}
where $\Phi_{\textrm{astro}}$ is the normalization for a single neutrino flavor and charge (i.e., either $\nu$ or $\bar{\nu}$) and $\gamma_{\textrm{astro}}$ is the spectral index.  
This analysis is not sensitive to the details of the spectral shape of the astrophysical neutrino flux. However, $\Phi_{\textrm{astro}}$ and $\gamma_{\textrm{astro}}$ are included as nuisance parameters with central values and correlated priors chosen according to a combined fit of the results of Refs.~\cite{IceCube:2020wum,IceCube:2016umi,IceCube:2020acn}.
The central values used are $\Phi_{\textrm{astro}} = \SI{0.787e-18}{\per\GeV\per\steradian\per\second\per\square\cm}$ and $\gamma_{\textrm{astro}} = 2.5$.
The analysis assumes equal flavor content (\nue:\numu:\nutau::1:1:1) and equal $\nu$/$\bar\nu$ flux. 

In a 3+1 model, the resonance occurs as the neutrinos propagate through the Earth's matter.
The effects of the parameters, \Umufsq, \Utaufsq, and \Dmqfo,  are calculated using the neutrino propagation software \texttt{nuSQuIDS}~\cite{Arguelles:2021twb}, which adopts the neutrino mixing matrix~\cite{Esteban:2020cvm}  and the interaction cross-section~\cite{Cooper-Sarkar:2011jtt} relevant at these energies to model coherent and incoherent interactions, including tau neutrino regeneration~\cite{Halzen:1998be}.

An example of the effect that the existence of a sterile neutrino would have on the oscillations was discussed earlier and shown in \Cref{fig:oscillograms} as a function of true neutrino energy $E_\nu$ and true neutrino direction \cosz. 
The Earth's density as a function of radius is taken to follow the PREM model~\cite{Dziewonski:1981xy}.

Neutrino interactions are detected using the IceCube Neutrino detector located at the geographic South Pole.   For a detailed description of the apparatus, see Ref. \cite{Aartsen:2016nxy}.   Upgoing charged-current (CC) muon \mbox{(anti)neutrino} interactions that occur below and within the IceCube detector are identified via muon tracks observed in the detector.  
Tracks are reconstructed~\cite{IceCube:2013dkx,IceCube:2015qii} based on emitted Cherenkov light collected by `Digital Optical Modules' (DOMs)~\cite{Abbasi:2008aa} arranged in two hexagonal arrays.
The main array comprises 78 strings, each carrying 60 DOMs with 17\,m spacing, separated by 125\,m.
The DeepCore~\cite{Collaboration:2011ym} subdetector, an 8-string array with variable string-spacing between \SI{42}{\m} and \SI{72}{\m} and vertical DOM separation of \SI{7}{\m}, is also used in this analysis.

This analysis follows recent TeV-scale IceCube 3+1 analyses ~\cite{IceCube:2020phf,IceCube:2020tka,IceCubeCollaboration:2022tso} in the event selection and reconstruction procedure. A total of 305,735 track-like events, selected to have reconstructed energy between \SI{500}{\GeV} and \SI{9976}{\GeV}, were collected over a live time of 7.634 years between May 2011 and May 2019.
Tracks are selected so their reconstructed direction is within $-1 \leq \cosz \leq 0$, where $\cosz=-1$ corresponds to upgoing tracks and $\cosz=0$ to events at the horizon.
Even at the horizon, the matter traversed corresponds to \SI{124}{\km} of water-equivalent overburden at the top of the detector.
Hence, upgoing tracks are likely to originate from muon-neutrino interactions near the detector rather than from atmospheric muons penetrating through the Earth.
The selected sample is therefore expected to be composed of >99.9\% \numu+\numubar CC events~\cite{IceCube:2020tka}.
Since the event selection includes events that originate outside of the detector as well as tracks that exit the detector, the muon energy resolution is $\sigma\left(\log_{10}(E_{\mu})\right) \sim 0.5$. However, the combination of long track lengths of $\mathcal{O} (\SI{1}{\km})$ and large detector size allows direction reconstruction to better than \ang{1}~\cite{IceCube:2013dkx,Weaver:2015bja}.
Of the events in the selection, we expect ${\sim}20\%$ of events to originate from the bedrock, which lies approximately \SI{362}{\meter} below the bottom of the detector. 

Simulated neutrino interaction points are chosen and weighted using the \texttt{LeptonInjector} and \texttt{LeptonWeighter} software~\cite{IceCube:2020tcq}, respectively. 
For different sterile neutrino models, \texttt{LeptonWeighter} reweights each simulated event by the oscillated neutrino flux outputted by \texttt{nuSQuIDS}.
After selecting a neutrino interaction point, the secondary products are propagated through the rock and ice using the \texttt{PROPOSAL} software~\cite{Koehne:2013gpa}.

The \nutau /\nutaubar CC interactions will create $\tau^{-}$/$\tau^{+}$ that can decay leptonically to $\mu^{-}$/$\mu^{+}$, producing tracks in the sample.
While the direct production of tau-neutrinos in the dominant atmospheric flux is very low, nonzero \Utaufsq , i.e. nonzero $\theta_{34}$ (Eq. \ref{Udefs}), allows for non-negligible $\numubar \rightarrow \nutaubar$ appearance above \SI{500}{\GeV}.   Tau neutrinos also contribute to the astrophysical flux.
Past matter-induced IceCube sterile neutrino analyses~\cite{IceCube:2020phf,IceCube:2020tka} did not simulate tau-flavor contributions.
For this analysis, \nutau /\nutaubar simulations were produced and 
\texttt{PROPOSAL} was modified to account for the $\tau$ polarization when decaying into a $\mu$.
A separate $\nutau$ interaction cross-section is not needed, as mass effects at energies far above the $\tau$-lepton mass are negligible. 
For the sterile hypothesis parameters used in \Cref{fig:oscillograms}, $\numubar \rightarrow \nutaubar$ appearance contributes an additional 2\% of tracks in the upward direction, while contributing negligible events from the horizon. 

The systematic uncertainties adopted in this analysis have been detailed in previous IceCube analyses~\cite{IceCube:2020phf,IceCube:2020tka,IceCubeCollaboration:2022tso}. 
The list appears in Table~\ref{tab:sys}. 
The eighteen parameters fall into four categories: conventional atmospheric flux parameters, detector parameters, astrophysical neutrino flux parameters, and cross-section parameters.
In the analysis, different values for the systematic parameters modify the weights for the simulated events.
No uncertainties were placed on the Standard Model neutrino oscillation parameters.

\begin{table}[t!]
\centering
\begin{tabular}{ l c c}
\hline
\hline
\textbf{Systematic Parameter} & Prior  & Pull($\sigma$)\\
\hline
\hline 
\multicolumn{3}{c}{\textbf{Conventional Flux Parameters}}\\
\hline
\hline
Normalization ($N$)   &   1.00 $\pm$ 0.40  & 0.51\\
\hline
Spectral shift ($\Delta\gamma_{\mathrm{conv.}}$)   &  0.00 $\pm$ 0.25  & 0.44\\
\hline
Atm. Density      &   0.00 $\pm$ 1.00   & -0.03\\
\hline
WP          & 0.00 $\pm$ 0.40  & -0.13\\
\hline
WM         & 0.00 $\pm$ 0.40  & -0.06\\
\hline
YP        & 0.00 $\pm$ 0.30 & -0.35\\
\hline
YM         &  0.00 $\pm$ 0.30 & -0.04\\
\hline
ZP          &0.00 $\pm$ 0.12  & -0.02\\
\hline
ZM            &  0.00 $\pm$ 0.12  & 0.01\\
\hline
\hline
\multicolumn{3}{c}{\textbf{Detector Parameters}}\\
\hline
\hline
DOM Efficiency    & 0.97 $\pm$ 0.10  & -0.02\\
\hline
Hole Ice (p$_2$)  & -1.00 $\pm$ 10.00 & -0.23\\
\hline
Ice Gradient 0   & 0.00 $\pm$ 1.00*  & 0.10\\
\hline
Ice Gradient 1    &  0.00 $\pm$ 1.00* & 0.09\\
\hline
\hline
\multicolumn{3}{c}{\textbf{Astrophysics Parameters}}\\
\hline
\hline
Normalization ($\Phi_{\mathrm{astro.}}$)     &  0.787 $\pm$ 0.36* & 0.99\\
\hline
Spectral shape ($\Delta\gamma_{\mathrm{astro.}}$)   & 2.50 $\pm$ 0.36* & 0.92\\
\hline
\hline
\multicolumn{3}{c}{\textbf{Cross Section Parameters}}\\
\hline
\hline
Cross Section $\sigma_{\nu_\mu}$  & 1.00 $\pm$ 0.03 & -0.00\\
\hline
Cross Section $\sigma_{\overline{\nu}_\mu}$    & 1.000 $\pm$ 0.075  & 0.02\\
\hline
Kaon Energy Loss $\sigma_{KA}$   &  0.00 $\pm$ 1.00 & -0.13\\
\hline
\hline
\end{tabular}
\caption{The list of systematic uncertainties used in this analysis. The second column provides their central values and prior widths, while the third column gives the best-fit value pulls from the central value. Asterisks indicate correlated priors. Atmospheric flux parameterizations involving WP through ZM are defined in Ref.~\cite{IceCube:2020tka}.}
\label{tab:sys}
\end{table}

The majority of these nuisance parameters concern the conventional atmospheric flux.
Two parameters quantify the uncertainty in the cosmic ray flux.
The first is an overall normalization factor $N$, while the second is a spectral shape correction term $\Delta \gamma_{\mathrm{conv.}}$ applied as 
\begin{equation}
    \Phi(E_\nu; \Delta \gamma_{\mathrm{conv.}}) = N~\Phi_{\mathrm{conv.}}(E_\nu) \left( \frac{E_\nu}{\SI{2.2}{\TeV}} \right)^{-\Delta \gamma_{\mathrm{conv.}}},
\end{equation}
where $\Phi_{\mathrm{conv.}}$ is the nominal conventional neutrino flux and $\Delta \gamma_{\mathrm{conv.}}$ is centered at zero.
The only significant change from previous analyses~\cite{IceCube:2020phf,IceCube:2020tka,IceCubeCollaboration:2022tso} was to extend the prior on the cosmic ray spectral slope width $\Delta \gamma_{\mathrm{conv.}}$ from 0.03 to 0.25, to accommodate deviations from the Hillas-Gaisser 2012 H3a cosmic ray model spectrum recently observed in the rigidity region of interest $E = \SI{e3}{\GeV} - \SI{e4}{\GeV}$~\cite{Lipari:2019jmk}.

Another six conventional flux nuisance parameters vary the meson production rate in the hadronic interactions between cosmic rays and the Earth's atmosphere.
The six parameters correspond to the uncertainties for kaon productions in different sectors of the interaction kinematic phase space, as described in Ref.~\cite{Barr:2006it}.
The six parameters are labeled in \Cref{tab:sys} as WP, WM, YP, YM, ZP, and ZM.
The first letter for each label (W, Y, Z) corresponds to the kinematic phase space where the uncertainty is applied (see Fig. 2 in Ref.~\cite{Barr:2006it}), and the second letter (M, P) defines whether that uncertainty is applied to the negatively (M) or positively (P) charged meson.
Of the remaining uncertainties in Ref.~\cite{Barr:2006it}, they either correspond to low energy interactions ($\SI{<30}{\GeV}$) or to pion interactions.
They are found to be negligible for this analysis, since the neutrino flux above $\mathcal{O}(\SI{100}{\GeV})$ is primarily produced through kaon decays \cite{IceCube:2020tka}.

A final conventional flux parameter quantifies uncertainties in the Earth's atmospheric density profile.

Detector uncertainties account for another four nuisance parameters.
The first uncertainty is the DOM efficiency, which  accounts for uncertainties both in the DOM response as well as uncertainties in bulk ice properties that uniformly affect the detection of photons throughout the detector.
The second uncertainty pertains to the refrozen ice immediately surrounding the DOM strings.
This parameter, labeled ``Hole Ice ($\mathrm{p}_{2}$)" in \Cref{tab:sys}, is more thoroughly described in Ref. \cite{IceCube:2020tka}.

Two further parameters are implemented for the uncertainties and non-uniformities of the bulk ice of the Antarctic glacier.
As described in Ref.~\cite{IceCube:2019lxi}, the depth dependence of the bulk ice property can be expressed in terms of a Fourier decomposition.
The effect of these Fourier modes can be parameterized by two basis functions, and the amplitudes of these basis functions are implemented as systematic parameters \cite{IceCube:2020tka}.
In \Cref{tab:sys}, they are labeled as ``Ice Gradient 0'' and ``Ice Gradient 1.''

Two nuisance parameters are included to vary the normalization and spectral shape of the astrophysical neutrino flux given in \Cref{eq:astroflux}.

Finally, three cross-section nuisance parameters are included. Two are for uncertainties in neutrino and antineutrino interactions, and one last parameter is for kaon interaction cross-section uncertainties as they propagate through the Earth's atmosphere. 

Plots showing how the expected event rate changes as a function of the systematic parameters are included in \ref{sec:sys}. 

Two analyses were conducted: a frequentist parameter estimation and a Bayesian model comparison. 
In both cases, the physics parameters are sampled in a three-dimensional grid in (\Dmqfo, \Umufsq, \Utaufsq)-space.
The parameter bounds and step sizes are:
\begin{align*}
\Dmqfo &\in [0.1,50]\ \mathrm{eV}^{2}\ \textrm{in steps of 0.1 in}  \log_{10}(\Dmqfo),\\
\Umufsq &\in [0.001, 0.5]\ \textrm{in steps of 0.1 in} \log_{10}(|U_{\mu 4}|^{2}),\\
\Utaufsq &\in [0.001, 0.5]\ \textrm{in steps of 0.2 in}  \log_{10}(|U_{\tau 4}|^{2}).
\end{align*}
The abstract and conclusion state the frequentist result.

The data are binned in two dimensions: reconstructed cosine of the zenith angle, \cosz, and $\log_{10}$ of the reconstructed muon energy, $E$.
The reconstructed zenith is bounded between $\cosz = -1$ and $\cosz = 0$, with 20 bins of width $\Delta \cosz = 0.05$.
The reconstructed muon energy is bounded between \SI{500}{\GeV} and \SI{9976}{\GeV}, in 13 bins of width $\Delta \log_{10}(E/\textrm{GeV}) = 0.1$.

A likelihood is constructed 
\begin{equation}
    \mathcal{L}(\vec{\theta}, \vec{\theta}_{\eta}) = \mathcal{L}_{\mathrm{eff}}(\vec{\theta}, \vec{\theta}_{\eta})
    \Pi(\vec{\theta}_{\eta}),
\end{equation}
where $\mathcal{L}(\vec{\theta}, \vec{\theta}_{\eta})$ is the likelihood as a function of the fitted physics parameters $\vec{\theta} = \{\Dmqfo, \Umufsq, \Utaufsq\}$ and systematic parameters $\vec{\theta}_{\eta}$.
$\mathcal{L}_{\mathrm{eff}}$ is a modified Poisson distribution to account for limited Monte Carlo statistics in estimating an expected event rate~\cite{Arguelles:2019izp}. 
Systematic parameters are incorporated by factoring in a prior likelihood for each parameter, $\Pi(\vec{\theta}_{\eta})$, assuming a normal distribution with a given central value and width.

For the frequentist analysis, the systematic uncertainties are accounted for by using the profile likelihood at each tested physics point 
\begin{equation}
     \mathcal{L}_{\mathrm{profile}}(\vec{\theta}) = \max_{\vec{\theta}_{\eta}} \mathcal{L}(\vec{\theta}, \vec{\theta}_{\eta})
    \Pi(\vec{\theta}_{\eta})
\end{equation}

To construct confidence regions, we use the test statistic
\begin{align}
    \mathrm{TS}({\vec{\theta}}) &= -2 \Delta \log \mathcal{L}_{\mathrm{profile}}(\vec{\theta}) \\
     &= -2 ( \log \mathcal{L}_{\mathrm{profile}} (\vec{\theta}) - \log \mathcal{L}_{\mathrm{profile}} (\hat{\vec{\theta}})  ), 
\end{align}
where $\hat{\vec{\theta}}$ is the physics parameter point that maximizes $ \mathcal{L}_{\mathrm{profile}}$. 

We also perform a Bayesian analysis, where we calculate the Bayesian `evidence' at each physics point 
\begin{equation}
    \mathcal{E}(\vec{\theta})=\int d\vec{\theta}_{\eta} \mathcal{L}(\vec{\theta}, \vec{\theta}_{\eta}) \Pi(\vec{\theta}_{\eta}),
    \label{eq:evidence}
\end{equation}
and compare it to the evidence of the null model.
The ratio between the evidence at some physics parameter point and the evidence at the null gives the Bayes factor 
\begin{equation}
    \mathcal{K}(\vec{\theta}) = \frac{\mathcal{E}(\vec{\theta}=\mathrm{Null})}{\mathcal{E}(\vec{\theta})},
    \label{eq:bayesfactor}
\end{equation}
which quantifies the preference for the null model over the alternative model. 
The integral in \Cref{eq:evidence} is calculated using the \texttt{MultiNest} algorithm \cite{Feroz:2013hea}.

\begin{figure}[t]
    \centering
    \includegraphics[width=.99\linewidth]{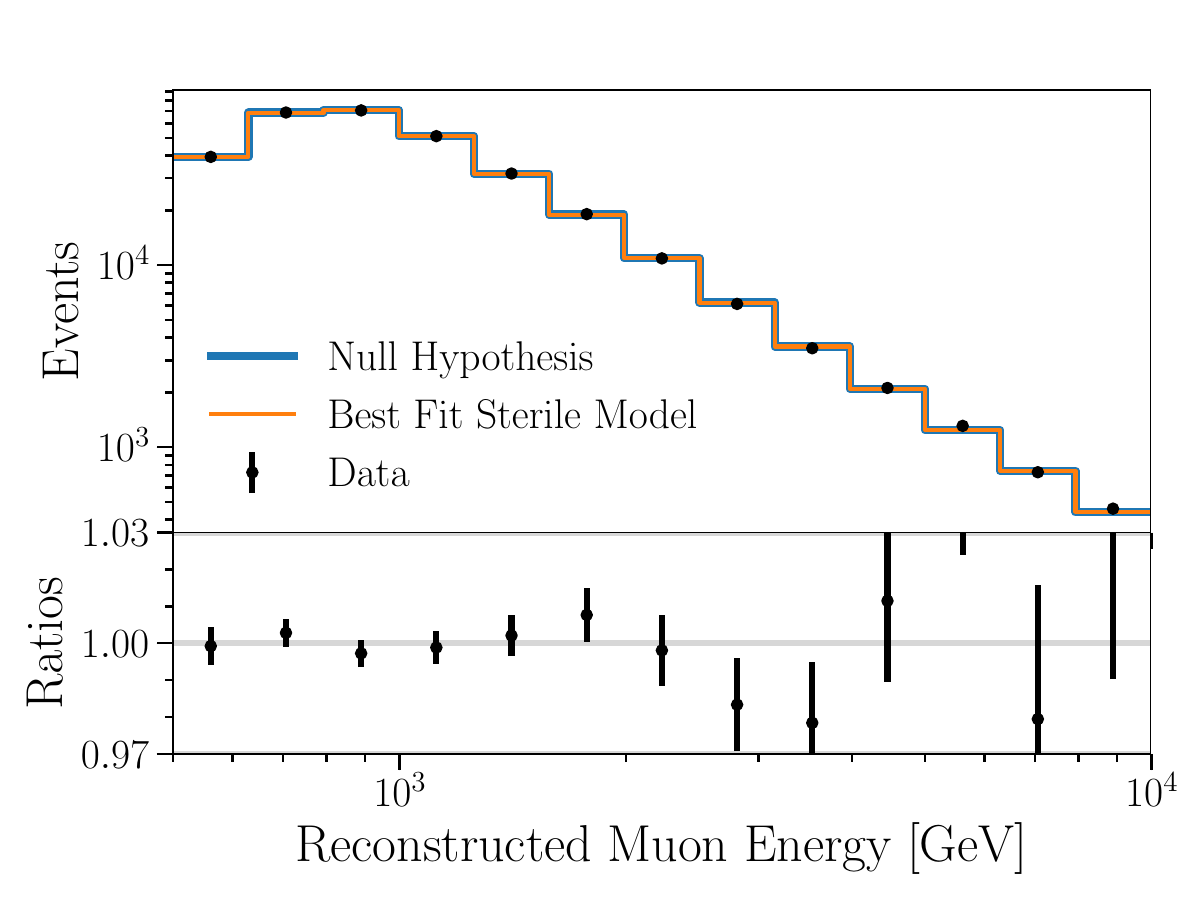}
    \includegraphics[width=.99\linewidth]{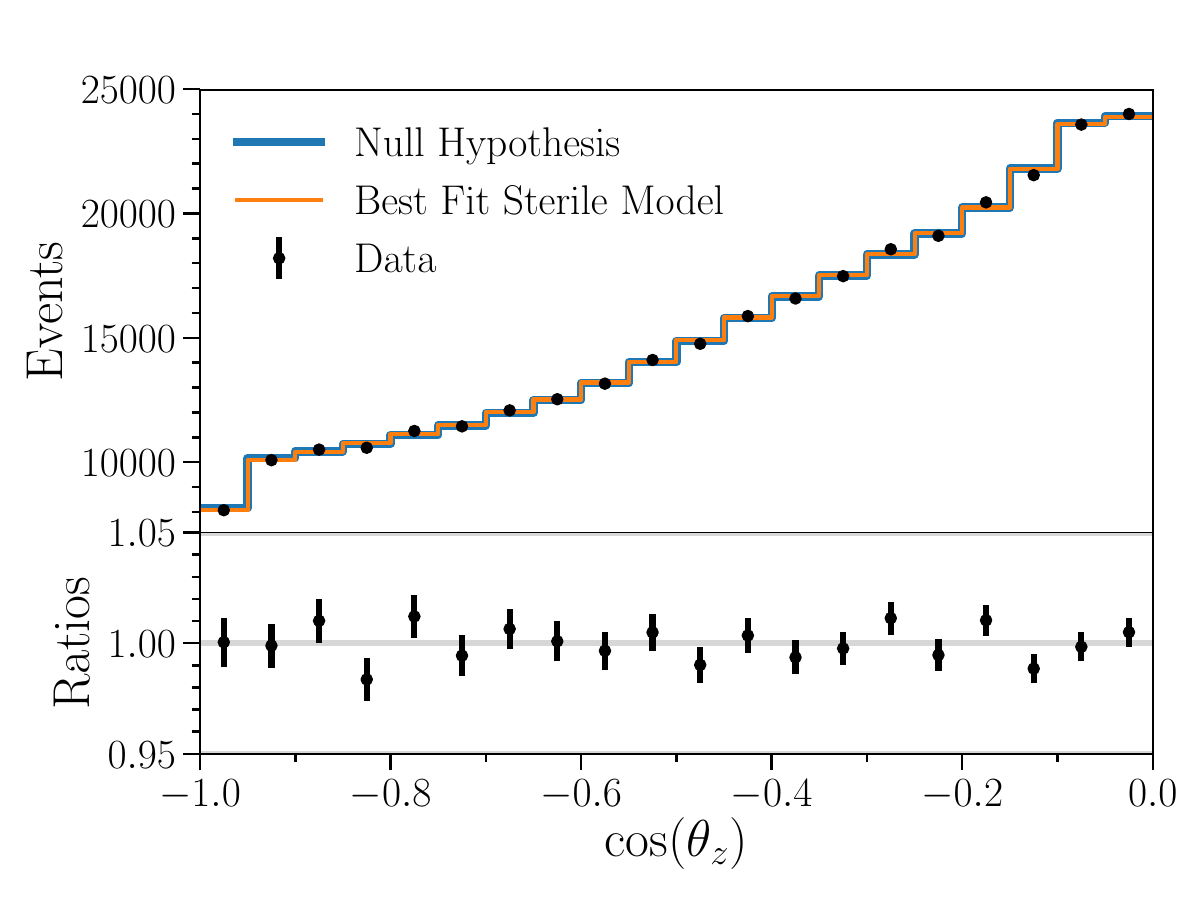}
    \caption{The 1D distribution of events selected as a function of reconstructed energy (top) and direction (bottom). Overlaid with the data are the expected distributions from the no-sterile null model (blue) and the best-fit sterile neutrino model (orange). The data points include error bars, which are too small to be observed. The ratio of the data to the best-fit model expectation is also presented.}
    \label{fig:onedim}
\end{figure}

To test our event selection and reconstruction, we compare our selected events with model expectations.
\Cref{fig:onedim} shows the observed events in reconstructed energy (top) and reconstructed direction (bottom) versus the no-sterile model,  shown in blue.
When the data are collapsed to a single dimension the analysis loses sensitivity to the expected sterile signal, therefore providing a test of the predicted event rate without biasing the physics results. 
\Cref{fig:onedim} shows, in orange, the expected distribution from the best-fit sterile model discussed in the next section.
As a goodness-of-fit test, when we run 2000 pseudo-experiments for this best-fit sterile model, we find that the 1D data distributions fall within expectation for the energy distribution with a p-value of 31\%, and for the zenith distribution with a p-value of 39\%.

\section{Results and discussion \label{sec:results}}

\begin{figure*}[tb!]
    \centering
    \includegraphics[width=.75\linewidth]{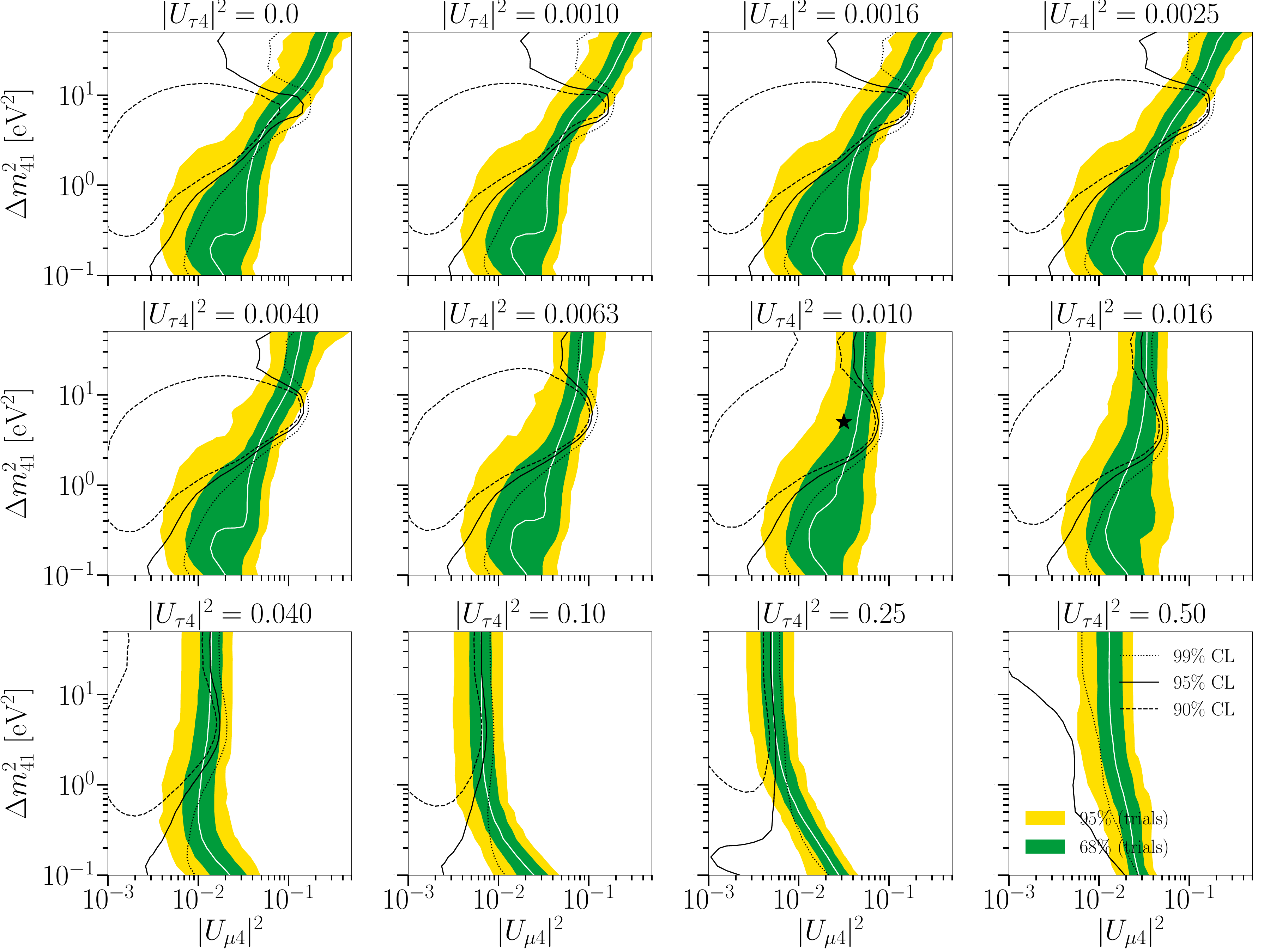}
    \caption{The 99\% (dotted lines), 95\% (solid lines), and 90\% (dashed lines) CL regions for a selection of \Utaufsq values tested, where the best-fit is marked with the black star~($\bigstar$). The contours are drawn assuming Wilks' theorem with three degrees of freedom as this is a three-dimensional fit. 
    The assumption of Wilks' theorem results in the null hypothesis being allowed within the 95\% CL. Pseudo-experiments, on the other hand, show that the sterile hypothesis is preferred over the null at 4.3\%. 
    The colored bands show where the middle 68.27\% (green) and 95.45\% (yellow) of simulated 99\% CLs for no signal lie.}
    \label{fig:frequentist}
\end{figure*}

For the frequentist analysis, the best-fit point is found at $\Dmqfo = \SI{5.0}{\eV\squared}$, $\Umufsq = 0.032$, and $\Utaufsq = 0.010$. 
The 99\%, 95\%, and 90\% confidence level (CL) regions of the fit are shown in \Cref{fig:frequentist}, assuming Wilks' theorem and three degrees of freedom for a subset of the \Utaufsq range sampled. 
In addition, the 99\% CL median sensitivity is shown, calculated using 300 pseudo-experiments assuming the no-sterile hypothesis.
The colored bands show where the middle 68.27\% and 95.45\% of the  simulated 99\% CLs lie. The expected $\numubar \to \numubar$ disappearance and $\numubar \to \nutaubar$ appearance oscillograms for the best-fit point can be found in \ref{sec:bfoscillogram}.

As a goodness-of-fit test, pseudo-experiments were run with the best-fit parameters injected as the true hypothesis.
For each of these pseudo-experiments, the likelihood at the injected point $\mathcal{L}_{\mathrm{profile}}(\vec{\theta}_\mathrm{injected})$ was calculated to obtain a distribution with which to compare the observed likelihood at the best-fit point. 
This comparison finds that the observed data are within expectation, with a p-value of 87\%. 
A similar exercise, with the null model as the injected and fitted parameter point, gives a p-value of 75\%. 

The best-fit nuisance parameters are all within $1\sigma$ of their associated nominal values, as listed in the third column of \Cref{tab:sys}.
This can be indicative of an overestimation of the systematic priors. 

\begin{figure*}[tb!]
    \centering
    \includegraphics[width=.99\linewidth]{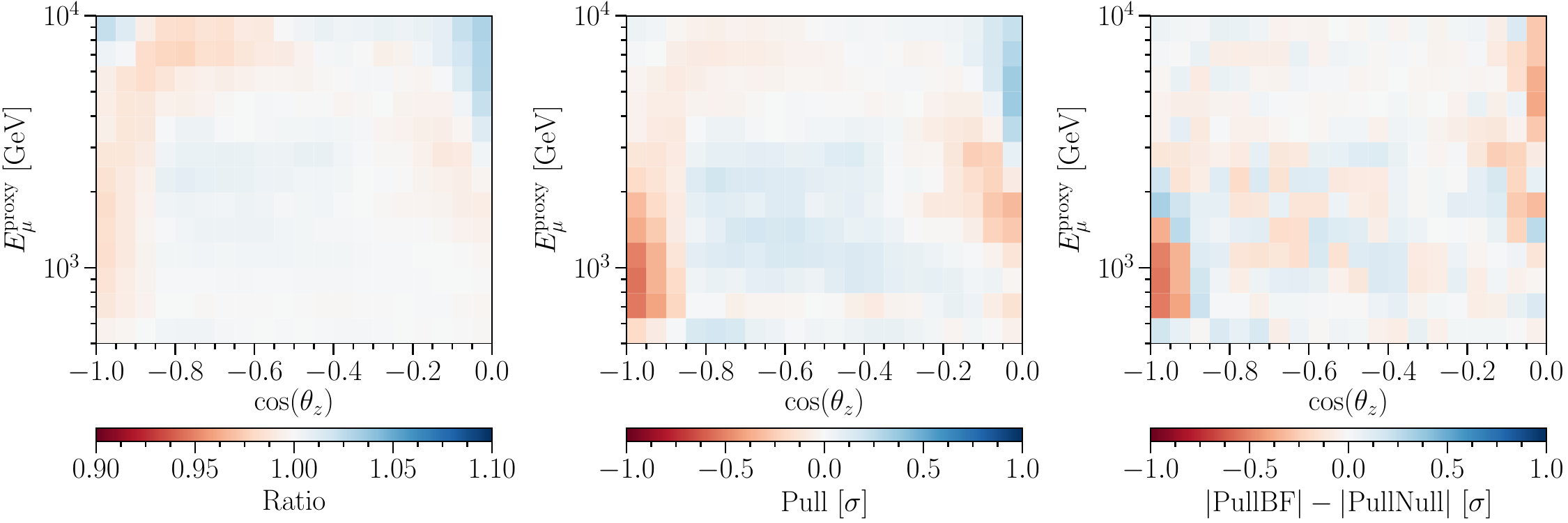}
    \caption{\textbf{Left:} Ratio of the expected event rate for the best-fit hypothesis over the null hypothesis, bin-by-bin, in the $\log_{10}(E)$ versus $\cos(\theta_z)$ plane. 
    \textbf{Middle:} Statistical pulls of the best-fit predicted event rate compared to the null expectation, bin-by-bin.
    \textbf{Right:} Difference between the absolute value of the pull of the data from the null hypothesis, and the absolute value of the pull of the data from the best-fit hypothesis.  
    Red bins indicate where the data are in better agreement with the best-fit hypothesis, than with the null hypothesis;
    these are the bins driving the fit to the best-fit hypothesis point.}
    \label{fig:datapulls}
\end{figure*}

\Cref{fig:datapulls} is helpful in understanding the best-fit 3+1 model.
The left plot shows the bin-by-bin ratios between the predicted event rates at the best-fit point and at the null hypothesis.
One can think of this prediction as the oscillogram smeared by uncertainties but without statistical variations expected in data.
The characteristic deficit at low $\cos{\theta_z}$ that one associates with the resonance in a non-zero $\Utaufsq$ model is evident.
An arc that results from the $L/E$ dependence of vacuum oscillations, which appears in all 3+1 models, can be observed at higher values of $\cos{\theta_{z}}$.

In the proceeding plots in \Cref{fig:datapulls}, we show pulls, where we define the pull of model $N$ relative to model $M$ as
\begin{equation}
\mathrm{Pull}(N; M) = \frac{E(N)-E(M)}{\sqrt{E(M)}},
\end{equation}
where $E(N)$ is the expected event rate for model $N$.
We can also measure the pull of the data $D$ relative to a model so that $E(D)$ in the above equation is the measured event rate.

The middle plot in \Cref{fig:datapulls} presents the predicted pull, bin-by-bin, between the null hypothesis and the best-fit hypothesis, emphasizing the bins that have the most statistical power in distinguishing the two.
The right plot shows the difference between the absolute values of the pull of the data from either hypothesis, i.e., $|\mathrm{Pull (data; best~fit)}| - |\mathrm{Pull(data; null)}|$.
If we think of $|\mathrm{Pull (data; N)}|$ as a measure of the disagreement of the data from model $N$, then $|\mathrm{Pull (data; best~fit)}| - |\mathrm{Pull(data; null)}|$  highlights (in red) which bins have better agreement with the best-fit model compared to the null model.
These bins are most strongly contributing to the test statistic $-2\Delta \log\LH$.

The test statistic found at the best-fit versus the no-sterile hypotheses point gives a $-2\Delta \log\LH = 6.9$.
According to Wilks' theorem, with three degrees of freedom, the sterile hypothesis is favored over the null hypothesis with a probability of 7.5\%.
However, pseudo-experiments can be conducted in order to calculate a more accurate probability without relying on Wilks' theorem.
With 1000 simulations, we obtain a preference for the sterile hypothesis over the null hypothesis of 4.3\%, corresponding to a $\num{2}\sigma$ significance, as reported in the abstract and conclusion.

We can compare this to the result of a traditional 3+1 fit that assumes $\Utaufsq=0$.
If we restrict $\Utaufsq = 0$, we find the best-fit to be $\Dmqfo = \SI{5.0}{\eV\squared}$ and $\Umufsq = 0.025$, with $-2\Delta \log\LH = 4.7$.
Assuming Wilks' theorem, this gives a consistency with the null hypothesis  of 10\%, in agreement with pseudo-experiment simulations, and corresponding to $1.7\sigma$ significance.

Accurately determining preferred regions without relying on Wilks' Theorem by using pseudo-experiments, as in \Cref{fig:frequentist}, would require hundreds of simulations to be run at each sampled physics parameter point; this is computationally unfeasible.
Instead, we ran simulations at a selection of parameter points, described in \ref{sec:Wilks}.  This study found that assuming Wilks' theorem with three degrees of freedom provides similar or more conservative exclusions at the 99\%, 95\% and 90\% CL.
To reduce the computational overheads, we draw these confidence regions assuming Wilks' theorem.

\begin{figure}[t]
    \centering
    \includegraphics[width=.99\linewidth]{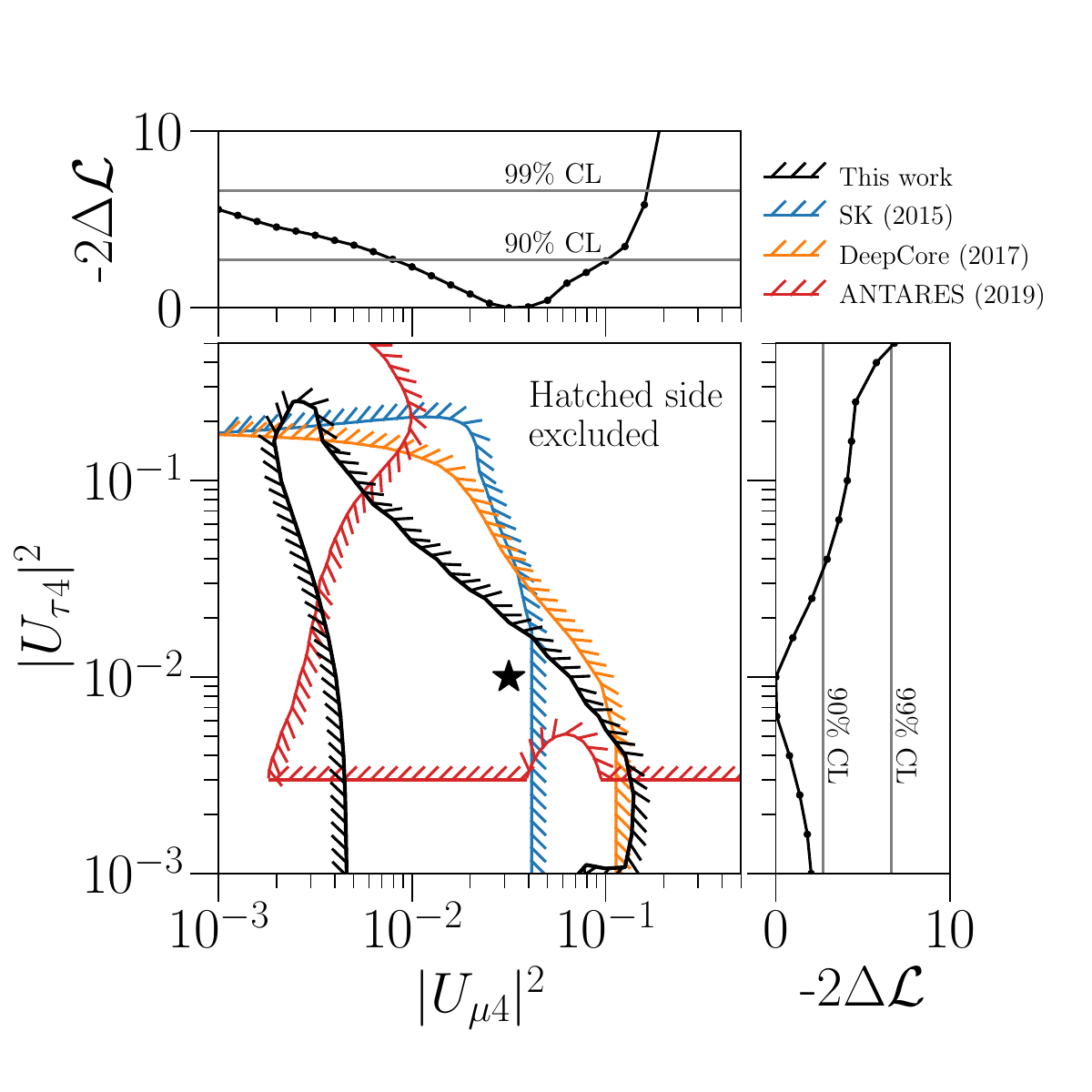}
    \caption{Comparison of the 90\% CL contour of this analysis to Super-Kamiokande (SK)~\cite{Super-Kamiokande:2014ndf}, IceCube-DeepCore (DeepCore)~\cite{IceCube:2017ivd}, and ANTARES~\cite{ANTARES:2018rtf}. Since other experiments do not perform a 3D fit over (\Dmqfo, \Umufsq, \Utaufsq), results are profiled over \Dmqfo to allow the comparison (bottom left).
    The hatched lines indicate the area which is excluded by the respective experiment. 
    For ANTARES, the horizontal line at $\Utaufsq = 0.003$ indicates the lower bound of that analysis.
    The top (right) plot shows our test statistic distribution when all parameters but \Umufsq (\Utaufsq) have been profiled over.  The relatively shallow $-2\Delta\log\mathcal{L}$ distributions indicate the best-fit point (star) is not strongly favored within the wide allowed range.
    The two gray lines mark the 90\% and 99\% critical values for Wilks' theorem in one dimension.}
    \label{fig:comparison}
\end{figure}

To compare to other experiments, we profile over \Dmqfo and compare contours in the (\Umufsq, \Utaufsq) parameter space.
We show these comparisons in \Cref{fig:comparison}.
The 90\% CL contours are shown in the lower left panel.
We see that, in this 2D space, the present analysis gives complementary coverage to other experiments.
The upper left and far right panels in \Cref{fig:comparison} show the $-2\Delta\log\mathcal{L}$ distribution when (\Dmqfo, \Utaufsq) and (\Dmqfo, \Umufsq), respectively,  have been minimized over.  
The grey lines show the $-2\Delta\log\mathcal{L}$ critical values for the 90\% and 99\% CL at one degree of freedom. 
For \Umufsq and \Utaufsq, the 90\% (99\%) CL ranges for our analysis are
\begin{align*}
    0.0081 < &\Umufsq < 0.10\ (\Umufsq < 0.16) \\
    & \Utaufsq < 0.035\ (\Utaufsq <0.48).
\end{align*}
  One sees that the $-2\Delta\log\mathcal{L}$ distribution is relatively shallow. Thus, the best-fit is not strongly preferred over a wide range of values in matrix element space. Therefore, while some matrix element combinations are disfavored by other experiments, other parameter combinations remain viable. 

\begin{figure*}
    \centering
    \includegraphics[width=.8\linewidth]{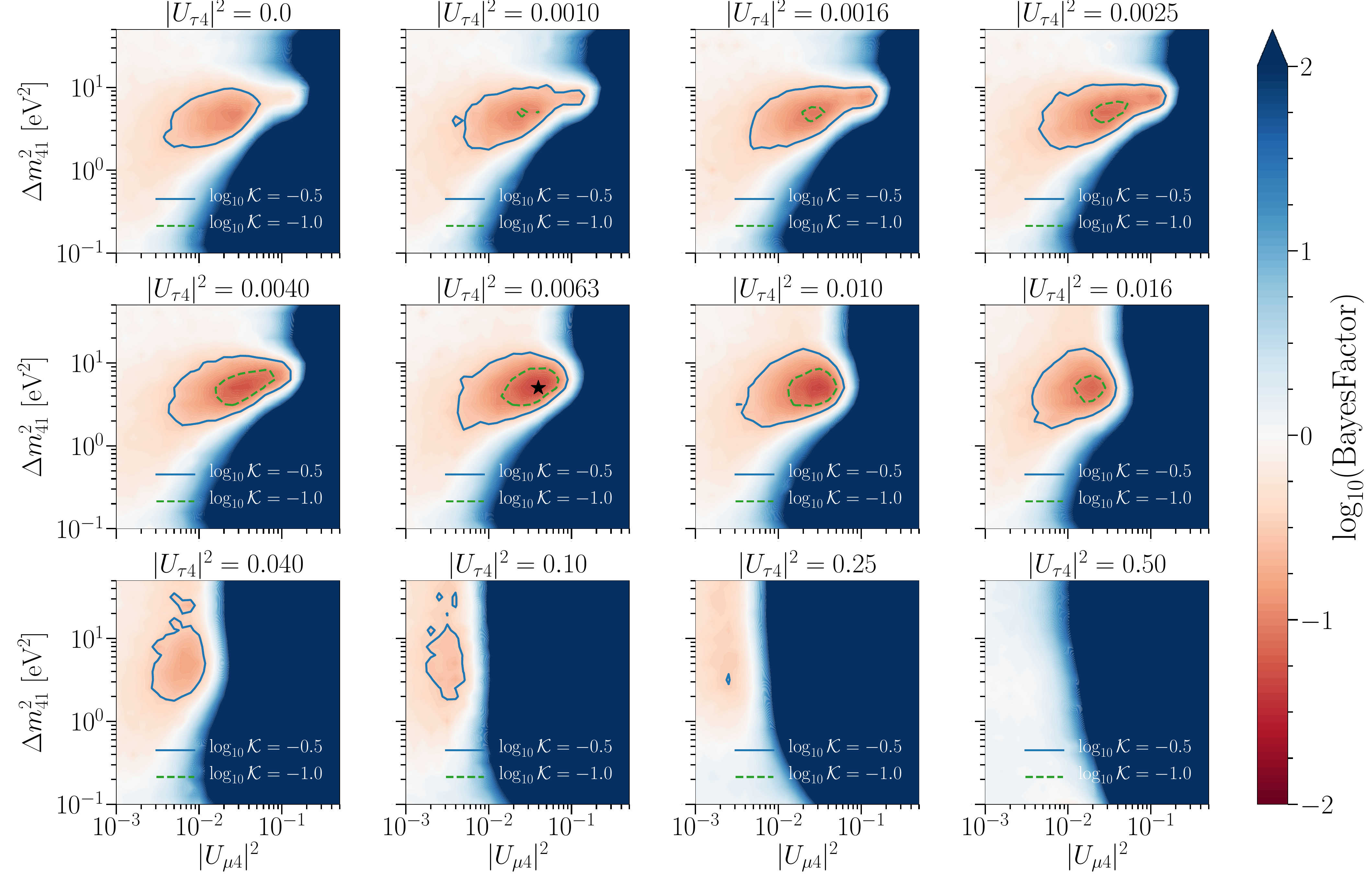}
    \caption{The Bayes Factor for a selection of \Utaufsq values tested. The Bayes Factor at each point is calculated with respect to the null model. The point with the most significant Bayes Factor is marked with the black star~($\bigstar$).}
    \label{fig:bayesian}
\end{figure*}

The best-fit Bayes factor, relative to the null, is $\log_{10}\left(\mathcal{K}\right) = -1.4$ at $\Dmqfo = \SI{5.0}{\eV\squared}$, $\Umufsq = 0.040$, and $\Utaufsq = 0.0063$.
With our sampling, this point is adjacent to the best-fit frequentist point.
Following Jeffreys' scale~\cite{Jeffreys:1939xee}, this corresponds to a ``strong'' preference over the null model. 
\Cref{fig:bayesian} shows the Bayesian fit across a subset of sampled values of \Utaufsq. The black star indicates the point with the greatest evidence. 

The Bayesian and frequentist fits have different interpretations. Bayesian credible regions indicate our expectation that the model parameters are at a given value. Frequentist confidence regions are constructed
to have a specified prior probability (the confidence level) of
including the true parameter values.
A summary comparing the results
of the Bayesian and frequentist fits can be found in \Cref{tab:results}.

\begin{table}[t]
\begin{center}
\begin{tabular}{ c c c  } 
Fit Param. & Frequentist & Bayesian   \\ 
\hline
$\Dmqfo$ & \SI{5.0}{\eV\squared} & \SI{5.0}{\eV\squared} \\
$\Umufsq$ & $0.032$ & $0.040$ \\
$\Utaufsq$ & $0.010$ & $0.0063$ \\  \hline
& $-2\Delta \log \LH = 6.9$ ($p=4.3\%$) & $\log_{10}\mathcal{K} = -1.4$
 \end{tabular}
\end{center}
\caption{Summary of the three-parameter-fit results. The results from the fequentist and Bayesian studies identify consistent parameter regions of interest.  \label{tab:results}}  
\end{table}

IceCube does not expect a substantial increase in statistical strength in the near future because another decade would be required to double the data set; nevertheless, future improvements to the analysis are possible. The
IceCube Upgrade~\cite{Ishihara:2019aao} that will come online in the next few years is focused on increasing detector efficiency at energies below that used in this analysis. While this upgrade will improve the understanding of the ice properties, the events it can detect will not lie in the energy range of interest for this specific analysis.
Therefore, near-future improvements to this study need to focus on the expansion of the data set, such as the inclusion of cascade event signatures~\cite{Smithers:2021orb} and improved event reconstruction in the relevant energy range by, for example, separating events that begin within the detector from events that start outside of the detector.

\section{Conclusion \label{sec:conclusion}}

This paper has presented the first sterile-neutrino analysis that fits simultaneously for \Dmqfo, \Umufsq, and \Utaufsq in a 3+1 sterile model.  
Our study uses 305,735 IceCube muon-track events, originating from $\nu_\mu$ and $\nu_{\tau}$ CC interactions, produced below and within the detector with energies between \SI{500}{\GeV} and \SI{9976}{\GeV}.   
The frequentist best-fit parameter point is at $\Dmqfo = \SI{5.0}{\eV\squared}$, $\Umufsq = 0.032$, $\Utaufsq = 0.01$, and preferred over the null hypothesis (three-neutrino) with a probability of 4.3\%.
The Bayesian result identifies preferred regions localized on parameter values consistent with the frequentist confidence limits, demonstrating a consistent picture.

As seen in \Cref{fig:comparison}, our result is consistent with other experiments when the analysis is reduced to two parameters, \Umufsq and \Utaufsq, through profiling in order to enable comparison.
However, information is lost when one reduces to these parameters because the full 3+1 model predicts dependence on \Dmqfo in the range of 1 to 10 eV$^2$, as can be seen in the three-parameter fits presented in \Cref{fig:frequentist}. 
Hence, experiments that by design average over this range of \Dmqfo lack essential sensitivity.
Our result, therefore, illustrates the importance of expanding to three parameters fits for \Dmqfo, \Umufsq, and \Utaufsq to test the 3+1 model, particularly for experiments that do not suffer fast oscillations for \Dmqfo in the 1--10 eV$^2$ range.

\section*{Acknowledgements}
The IceCube collaboration acknowledges the significant
contributions to this manuscript from the Massachusetts
Institute of Technology, Harvard University, and University of
Delaware groups.
The authors gratefully acknowledge the support from the following agencies and institutions:
USA {\textendash} U.S. National Science Foundation-Office of Polar Programs,
U.S. National Science Foundation-Physics Division,
U.S. National Science Foundation-EPSCoR,
U.S. National Science Foundation-Office of Advanced Cyberinfrastructure,
Wisconsin Alumni Research Foundation,
Center for High Throughput Computing (CHTC) at the University of Wisconsin{\textendash}Madison,
Open Science Grid (OSG),
Partnership to Advance Throughput Computing (PATh),
Advanced Cyberinfrastructure Coordination Ecosystem: Services {\&} Support (ACCESS),
Frontera computing project at the Texas Advanced Computing Center,
U.S. Department of Energy-National Energy Research Scientific Computing Center,
Particle astrophysics research computing center at the University of Maryland,
Institute for Cyber-Enabled Research at Michigan State University,
Astroparticle physics computational facility at Marquette University,
NVIDIA Corporation,
and Google Cloud Platform;
Belgium {\textendash} Funds for Scientific Research (FRS-FNRS and FWO),
FWO Odysseus and Big Science programmes,
and Belgian Federal Science Policy Office (Belspo);
Germany {\textendash} Bundesministerium f{\"u}r Bildung und Forschung (BMBF),
Deutsche Forschungsgemeinschaft (DFG),
Helmholtz Alliance for Astroparticle Physics (HAP),
Initiative and Networking Fund of the Helmholtz Association,
Deutsches Elektronen Synchrotron (DESY),
and High Performance Computing cluster of the RWTH Aachen;
Sweden {\textendash} Swedish Research Council,
Swedish Polar Research Secretariat,
Swedish National Infrastructure for Computing (SNIC),
and Knut and Alice Wallenberg Foundation;
European Union {\textendash} EGI Advanced Computing for research;
Australia {\textendash} Australian Research Council;
Canada {\textendash} Natural Sciences and Engineering Research Council of Canada,
Calcul Qu{\'e}bec, Compute Ontario, Canada Foundation for Innovation, WestGrid, and Digital Research Alliance of Canada;
Denmark {\textendash} Villum Fonden, Carlsberg Foundation, and European Commission;
New Zealand {\textendash} Marsden Fund;
Japan {\textendash} Japan Society for Promotion of Science (JSPS)
and Institute for Global Prominent Research (IGPR) of Chiba University;
Korea {\textendash} National Research Foundation of Korea (NRF);
Switzerland {\textendash} Swiss National Science Foundation (SNSF).

\bibliographystyle{elsarticle-num}
\bibliography{main}

\appendix

\section{Systematic parameter effects on event rate}\label{sec:sys}

\begin{figure*}
    \centering
    \includegraphics[width=0.99\linewidth]{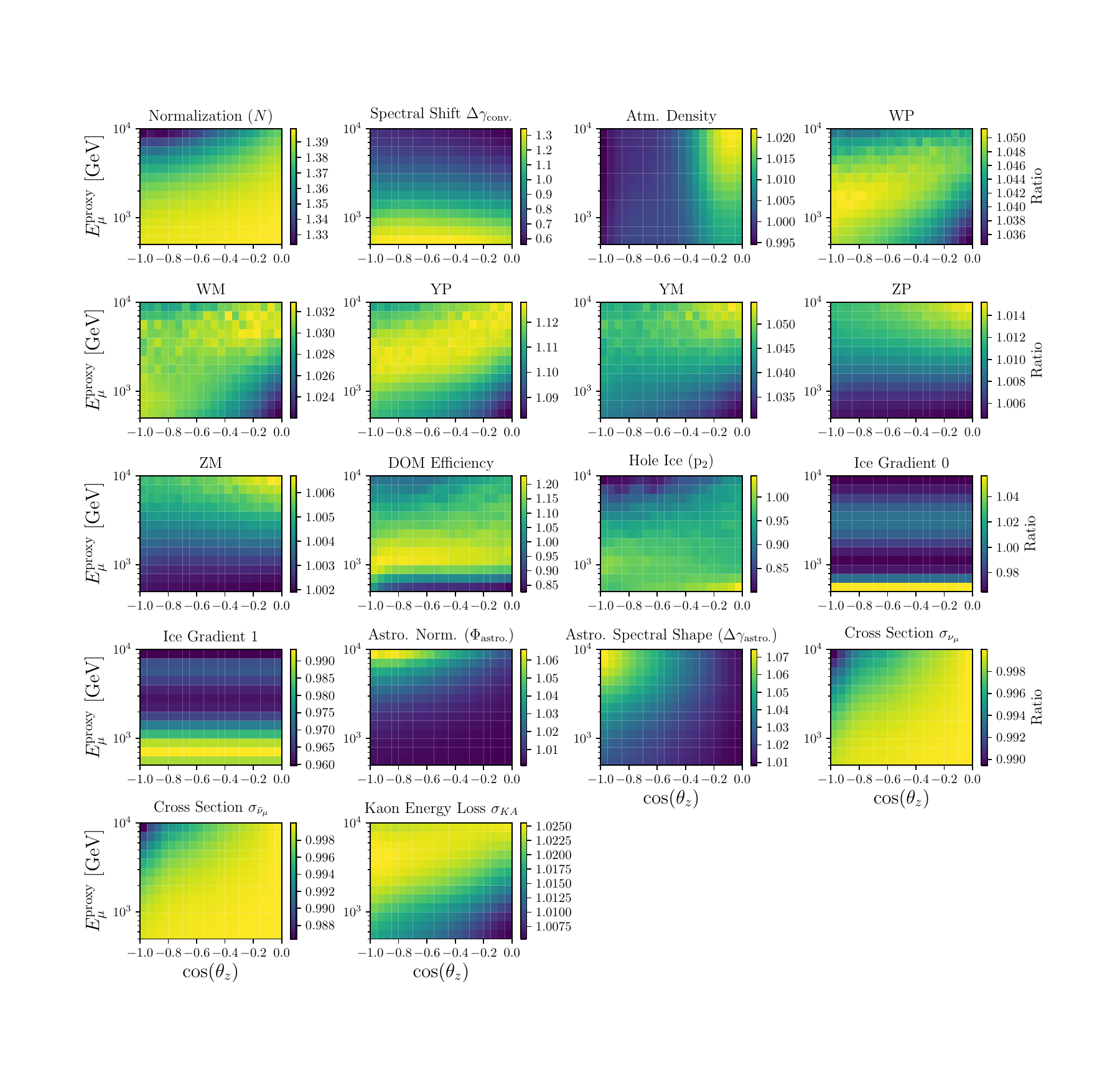}
    \caption{Each plot shows the ratio of the expected event rate when each systematic parameter is shifted up by $1 \sigma$ over the expected event rate with nominal systematic parameter values.
    All plots are made assuming the no sterile neutrino model.}
    \label{fig:sysshift}
\end{figure*}

\Cref{fig:sysshift} shows how the expected reconstructed event rate changes when the given systematic parameter is shifted up by $1\sigma$.

\section{Oscillogram at best-fit point}\label{sec:bfoscillogram}

\begin{figure}[t!]
    \centering
    \includegraphics[width=0.99\linewidth]{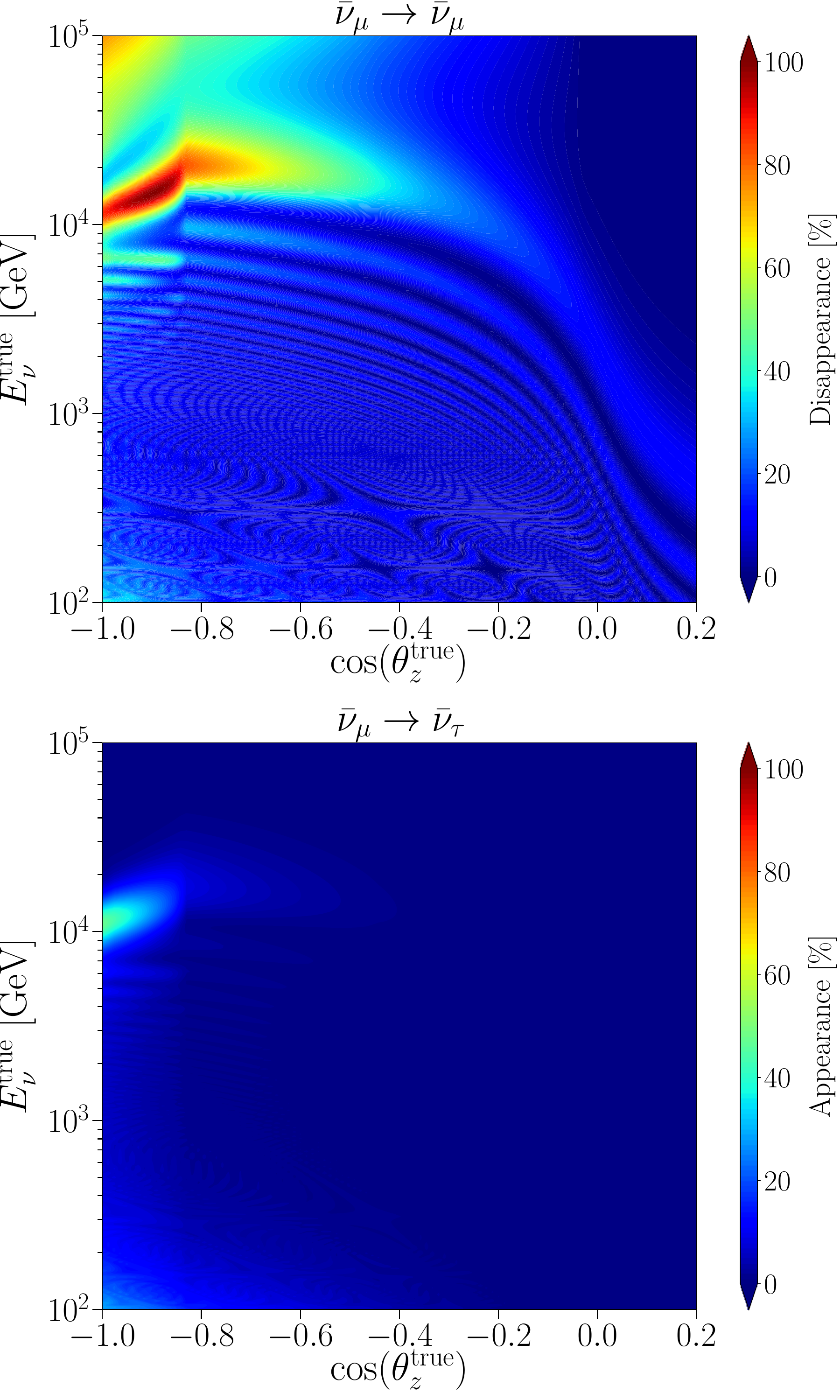}
    \caption{\textbf{Top:} The expected $\numubar \to \numubar$ disappearance oscillations at the best-fit point of $\Dmqfo = \SI{5.0}{\eV\squared}$, $\Umufsq = 0.032$, and $\Utaufsq = 0.010$.
    \textbf{Bottom:} The expected $\numubar \to \nutaubar$ appearance at the same best-fit point.}
    \label{fig:BFoscillogram}
\end{figure}

\Cref{fig:BFoscillogram} shows the oscillograms for $\numubar \to \numubar$ disappearance and $\numubar \to \nutaubar$ appearance at the best-fit point of $\Dmqfo = \SI{5.0}{\eV\squared}$, $\Umufsq = 0.032$, and $\Utaufsq = 0.010$. 

\section{Wilks' vs. frequentist coverage tests}\label{sec:Wilks}

The analysis presents confidence limits based on Wilks' theorem with three degrees of freedom.
In order to test the validity of this assumption,  frequentist "spot-checks" were performed.
In this study,
simulations were fit to obtain a sampled $-2\Delta\log\mathcal{L}$ distribution.
We report the results in \Cref{tab:spot}, where each column gives the true coverage found through simulations, which should be compared to the coverage at the defined Wilks' theorem confidence levels of 90\%, 95\% and 99\%.

\begin{table}[H]
\begin{center}
\begin{tabular}{ c | c c c } 
Sampled Point &  \multicolumn{3}{c}{Coverage, assuming Wilks' CL}  \\ 
  $(\Dmqfo, \Umufsq, \Utaufsq)$ & 90\% & 95\% & 99\% \\
\hline
$(0,0,0)$ & $93.9^{+0.8}_{-0.8}\%$ & $97.7^{+0.5}_{-0.6}\%$ & $99.4^{+0.2}_{-0.4}\%$ \\[1mm]
$(5,0.03,0.01)$ & $89.0^{+1.4}_{-1.6}\%$ & $95.2^{+1.0}_{-1.2}\%$ & $99.6^{+0.3}_{-0.5}\%$\\[1mm]
$$(1.0, 0.0063, 0.10)$$  & $93.4^{+1.1}_{-1.3}\%$ & $97.4^{+0.7}_{-0.9}\%$ & $99.2^{+0.4}_{-0.6}\%$\\[1mm]
$(10, 0.16, 0.0016)$ & $92.4^{+1.2}_{-1.4}\%$ & $96.8^{+0.8}_{-1.0}\%$ & $99.6^{+0.3}_{-0.5}\%$ \\
 \end{tabular}
\end{center}
\caption{Summary of the Wilk's coverage test at a selection of sampled points. \label{tab:spot}}  
\end{table}

\end{document}